\documentclass[letterpaper]{article}
\usepackage{aaai19}
\usepackage{times}
\usepackage{helvet,makecell}
\usepackage{courier}
\usepackage{graphicx,pifont,multirow,diagbox,array}
\usepackage{color}
\usepackage{float,amsmath}

\usepackage{subfigure}

\newcommand{\cmark}{\ding{51}}%
\begin{document}
\title{ \texttt{FakeNewsNet}: A Data Repository with News Content, Social Context and Spatiotemporal Information for Studying Fake News on Social Media}

\author{Kai Shu$^{1}$, Deepak Mahudeswaran$^{1}$, Suhang Wang$^{2}$, Dongwon Lee$^{2}$ and  Huan Liu$^{1}$ \vspace{0.1in} \\
$^{1}$Computer Science and Engineering, Arizona State University, Tempe, 85281, USA\\
$^{2}$College of Information Sciences and Technology, Penn State University, University Park, PA, 16802, USA\\
\{kai.shu, dmahudes, huan.liu\}@asu.edu, \{szw494, dongwon\}@psu.edu
}

\maketitle
\begin{abstract}
Social media has become a popular means for people to consume and share the news. At the same time, however, it has also enabled the wide dissemination of \textit{fake news}, i.e., news with intentionally false information, causing significant negative effects on society. To mitigate this problem, the research of  fake news detection has recently received a lot of attention. 
Despite several existing computational solutions on the  detection of fake news, however, the lack of comprehensive and community-driven fake news datasets has become one of major roadblocks. Not only existing datasets are scarce, they do not contain a myriad of features often required in the study 
%However, such a problem is non-trivial to solve,  often requiring multi-source information 
such as \textit{news content}, \textit{social context}, and \textit{spatiotemporal information}. 
%First, fake news is written to mislead people, which makes it difficult to detect fake news simply based on news contents. In addition to news contents, we need to explore social contexts such as user engagements and social behaviors. For example, a credible user's comment that ``this is fake news'' is a strong signal for detecting fake news. Second, spatiotemporal information such as how real or fake  news propagate and how users' opinions toward news pieces are changing are very important for extracting useful patterns for (early) fake news detection and intervention. 
%Thus, comprehensive datasets which contain \textit{news content}, \textit{social context}, and \textit{spatiotemporal information} could facilitate the study of fake news propagation, detection, and mitigation; while to the best of our knowledge, existing datasets only contains one or two aspects. 
Therefore, in this paper, to facilitate fake news related research, we present  a fake news data repository {\em FakeNewsNet}, which contains two comprehensive datasets with diverse features in \textit{news content}, \textit{social context}, and \textit{spatiotemporal information}. We present a comprehensive description of the FakeNewsNet, demonstrate an exploratory analysis of two datasets from different perspectives, and discuss the benefits of the FakeNewsNet for potential applications on fake news study on social media.

%The dataset collected has a rich set of linguistic as well as social features like second order engagements including likes, retweets and replies that could effective for fake news detection problem. 
%This dataset could pave way for multiple research direction in fake news detection.
\end{abstract}

\section{Introduction}
% {\color{blue} Do you want to call it a dataset or two datasets or a data repository which contains two datasets?}
% \begin{itemize}
% \item \textit{What is fake news and Why fake news detection is important?}
% {\color{blue}  Revise introduction according to abstract}
Social media has become a primary source of news consumption nowadays. Social media is cost-free, easy to access, and can fast disseminate posts. Hence, it acts as an excellent way for individuals to post and/or consume information. For example, the time individuals spend on social media is continually increasing\footnote{https://www.socialmediatoday.com/marketing/how-much-time-do-people-spend-social-media-infographic}. As another example, studies from Pew Research Center shows that around 68\% of Americans get some of their news on social media in 2018\footnote{http://www.journalism.org/2018/09/10/news-use-across-social-media-platforms-2018/} and this has shown a constant increase since 2016. 
Since there is no regulatory authority on social media, the quality of news pieces spread in social media is often lower than traditional news sources. In other words, social media also enables the widespread of fake news. Fake news \cite{shu2017fake} means the false information that is spread deliberately to deceive people. Fake news affects the individuals as well as society as a whole. First, fake news can disturb the authenticity balance of the news ecosystem. Second, fake news persuades consumers to accept false or biased stories. For example, some individuals and organizations spread fake news in social media for financial and political gains~\cite{shu2017fake}. It is also reported that fake news has an influence on the 2016 US presidential elections\footnote{https://www.independent.co.uk/life-style/gadgets-and-tech/news/tumblr-russian-hacking-us-presidential-election-fake-news-internet-research-agency-propaganda-bots-a8274321.html}. Finally, fake news may cause significant effects on real-world events. For example, ``Pizzagate'', a piece of fake news from Reddit, leads to a real shooting~\footnote{https://www.rollingstone.com/politics/politics-news/anatomy-of-a-fake-news-scandal-125877/}. Thus, fake news detection is a critical issue that needs to be addressed.

%{\color{blue} replace one of the reasons by ``impacting stock market\footnote{https://www.forbes.com/sites/kenrapoza/2017/02/26/can-fake-news-impact-the-stock-market/\#10b2838a2fac}''. It's not good to have these contents appear in the majority of your papers} 

Detecting fake news on social media presents unique challenges. First, fake news pieces are intentionally written to mislead consumers, which makes it not satisfactory to spot fake news from news content itself. Thus, we need to explore information in addition to news content, such as user engagements and social behaviors of users on social media. For example, a credible user's comment that ``This is fake news'' is a strong signal that the news may be fake. Second,  the research community lacks datasets which contain spatiotemporal information to understand how fake news propagates over time in different regions, how users react to fake news, and how we can extract useful temporal patterns for (early) fake news detection and intervention. Thus, it is necessary to have comprehensive datasets that have news content, social context and spatiotemporal information to facilitate fake news research. However, to the best of our knowledge, existing datasets only cover one or two aspects.  

Therefore, in this paper, we construct and publicize a multi-dimensional data repository \textit{FakeNewsNet}\footnote{https://github.com/KaiDMML/FakeNewsNet}, which currently contains two datasets with news content, social context, and spatiotemporal information.
The dataset is constructed using an end-to-end system,FakeNewsTracker\footnote{http://blogtrackers.fulton.asu.edu:3000}     \cite{shu2018fakenewstracker}.The constructed FakeNewsNet repository has the potential to boost the study of various open research problems related to fake news study. First, the rich set of features in the datasets provides an opportunity to experiment with different approaches for fake new detection, understand the diffusion of fake news in social network and intervene in it. Second, the temporal information enables the study of early fake news detection by generating synthetic user engagements from historical temporal user engagement patterns in the dataset~\cite{qian2018neural}. Third, we can investigate the fake news diffusion process by identifying provenances, persuaders, and developing better fake news intervention strategies~\cite{fakebookchapter}.  Our data repository can serve as a starting point for many exploratory studies for fake news, and provide a better, shared insight into disinformation tactics. We aim to continuously update this data repository, expand it with new sources and features, as well as maintain completeness. The main contributions of the paper are:
\begin{itemize}
	\item We construct and publicize a multi-dimensional data repository for various facilitating fake news detection related researches such as fake news detection, evolution, and mitigation;
    \item We conduct an exploratory analysis of the datasets from different perspectives to demonstrate the quality of the datasets, understand their characteristics and provide baselines for future fake news detection; and 
   	\item We discuss benefits and provides insight for potential fake news studies on social media with FakeNewsNet.
\end{itemize}

% The rest of the paper is organized as follows. We describe the background of fake news problem and the existing related datasets in Section~\ref{sec:related}. In Section~\ref{sec:data}, we introduce the details of data integration. We further perform exploration analysis on FakeNewsNet and provide some insights from different perspectives to study fake news in Section~\ref{sec:analysis}. In Section~\ref{sec:application}, we discuss the potential applications of the data repository, and we conclude with future work in Section~\ref{sec:conclude}.

% {\color{blue} lack comprehensive benchmark datasets, explain what comprehensive means, why need comprehensive}
%{\color{blue} explain more on challenges of fake news detection, a general idea is like this: (i) written to fool people, text only cannot give satisfactory results; require information in addition to news contents such as user comments or social behaviors; (ii) lack datasets with spatiotemporal information to understand how fake news propagates, how users react to fake news and how can we extract useful patterns from these for (early) fake news detection. Therefore, we ...   In other words, we list the challenges of (early) fake news detection, propagation, intervention and state the requirements of to solve these challenges, which leads the integration of a comprehensive datasets that has these features} %One major reason for this is the lack of comprehensive dataset which include wide range of news content and social features.

\section{Background and Related Work}\label{sec:related}
Fake news detection in social media aims to extract useful features and build effective models from existing social media datasets for detecting fake news in the future. Thus, a comprehensive and large-scale dataset with multi-dimension information in online fake news ecosystem is important. The multi-dimension information not only provides more signals for detecting fake news but can also be used for researches such as understanding fake news propagation and fake news intervention. Though there exist several datasets for fake news detection, the majority of them only contains linguistic features. Few of them contains both linguistic and social context features. To facilitate research on fake news, we provide a data repository which includes not only news contents and social contents, but also spatiotemporal information. For a better comparison of the differences, we list existing popular fake news detection datasets below and compare them with the FakeNewsNet repository in Table~\ref{tab:compare}.

%To provide comprehensive datasetThe dataset repository in this paper contains news content, social context Several datasets exist for the fake news study and detection. The existing datasets generally include linguistic features and social context features. Most of the fake news datasets are generally formed by annotation of the news articles by expert journalists, fact checking websites, industry detectors and crowd-sourced workers. Some of the publicly available benchmark datasets are listed as follows:

% \item \textit{What are existing datasets for fake news detection, and what are the limitations of them (check the survey paper, and search if there are other datasets needed to compare) compared with our dataset?}

\textit{\textbf{BuzzFeedNews}~\footnote{https://github.com/BuzzFeedNews/2016-10-facebook-fact-check/tree/master/data}}: This dataset comprises a complete sample of news published in Facebook from 9 news agencies over a week close to the 2016 U.S. election from September 19 to 23 and September 26 and 27. Every post and the linked article were fact-checked claim-by-claim by 5 BuzzFeed journalists. It contains 1,627 articles –826 mainstream, 356 left-wing, and 545 right-wing articles.

\textit{\textbf{LIAR}~\footnote{https://www.cs.ucsb.edu/~william/software.html}}: This dataset \cite{wang2017liar} is collected from fact-checking website PolitiFact. It has 12.8 K human labeled short statements collected from PolitiFact and the statements are labeled into six categories ranging from completely false to completely true as pants on fire, false, barely-true, half-true, mostly true, and true.

\textit{\textbf{BS Detector}~\footnote{https://github.com/bs-detector/bs-detector}}: This dataset is collected from a browser extension called BS detector developed for checking news veracity. It searches all links on a given web page
for references to unreliable sources by checking
against a manually compiled list of domains. The labels
are the outputs of the BS detector, rather than human
annotators.

\textit{\textbf{CREDBANK}~\footnote{http://compsocial.github.io/CREDBANK-data/}}: This is a large-scale crowd-sourced dataset \cite{mitra2015credbank} of around 60 million tweets that cover 96 days starting from Oct. 2015. The tweets are related to over 1,000 news events. Each event  is assessed for credibilities by 30 annotators from Amazon Mechanical Turk.

\textit{\textbf{BuzzFace}~\footnote{https://github.com/gsantia/BuzzFace}}: This dataset \cite{santia2018buzzface} is collected by extending the BuzzFeed dataset with comments related to news articles on Facebook. The dataset contains 2263 news articles and 1.6 million comments discussing news content. 

\textit{\textbf{FacebookHoax}~\footnote{https://github.com/gabll/some-like-it-hoax}}: This dataset \cite{tacchini2017some} comprises information  related to posts from the facebook pages related to scientific news (non- hoax) and conspiracy pages (hoax) collected using Facebook Graph API. The dataset contains 15,500 posts from 32 pages (14 conspiracy and 18 scientific) with more than 2,300,000 likes.

% \item \textit{Provide a table to list existing datasets, the properties of these datasets.}
From Table~\ref{tab:compare}, we observe that no existing public dataset can provide all possible features of news content, social context, and spatiotemporal information. Existing datasets have some limitations that we try to address in our data repository. For example, BuzzFeedNews only contains headlines and text for each news piece and
covers news articles from very few news agencies. LIAR dataset contains mostly short statements instead of entire news articles with the meta attributes. BS Detector data is collected and annotated by using a developed news veracity checking tool, rather than using human expert annotators. CREDBANK dataset was originally collected for evaluating tweet credibilities and the tweets in the dataset are not related to the fake news articles and hence cannot be effectively used for fake news detection. BuzzFace dataset has basic news contents and social context information but it does not capture the temporal information. The FacebookHoax dataset consists very few instances about the conspiracy theories and scientific news.

To address the disadvantages of existing fake news detection
datasets, the proposed FakeNewsNet repository collects multi-dimension information from news content, social context, and spatiotemporal information from different types of news domains such as political and entertainment sources.

\begin{table*}[!h]
\begin{center}
\caption{Comparison with existing fake news detection datasets}
\label{tab:compare}
\begin{tabular}{| l | c | c | c | c| c | c|c|c|}
\hline
\multirow{2}{*}{\diagbox{\textbf{Dataset}}{\textbf{Features}}} & \multicolumn{2}{ c |}{\textbf{News Content}} & \multicolumn{4}{ c |}{\textbf{Social Context}} & \multicolumn{2}{ c |}{\textbf{Spatiotemporal Information}}  \\ 
\cline{2-9}
& \textit{Linguistic} & \textit{Visual} & \textit{User} & \textit{Post} & \textit{Response} & \textit{Network} & \textit{Spatial} & \textit{Temporal} \\
\hline
BuzzFeedNews & \cmark  & &   &  &  &  & &\\ \hline
LIAR & \cmark &  &  &   &   &  &  &\\ \hline
BS Detector &  \cmark  &  &   &  &  &  &  &\\ \hline
CREDBANK & \cmark &  & \cmark& \cmark &  &  &\cmark  &\cmark \\ \hline
BuzzFace & \cmark  &  &   &\cmark  & \cmark & &  &\cmark\\ \hline
FacebookHoax & \cmark  &  & \cmark  &\cmark  & \cmark &  & & \\ \hline
\textbf{FakeNewsNet} & \cmark & \cmark & \cmark & \cmark & \cmark & \cmark & \cmark & \cmark \\ \hline
\end{tabular}
\end{center}
\end{table*}
% {\color{blue} second order is not a good word}
% \end{itemize}

\section{Dataset Integration}\label{sec:data}
\begin{figure}[!htp]
\centering{
\hspace{-0.2cm}
\includegraphics[scale=0.175]{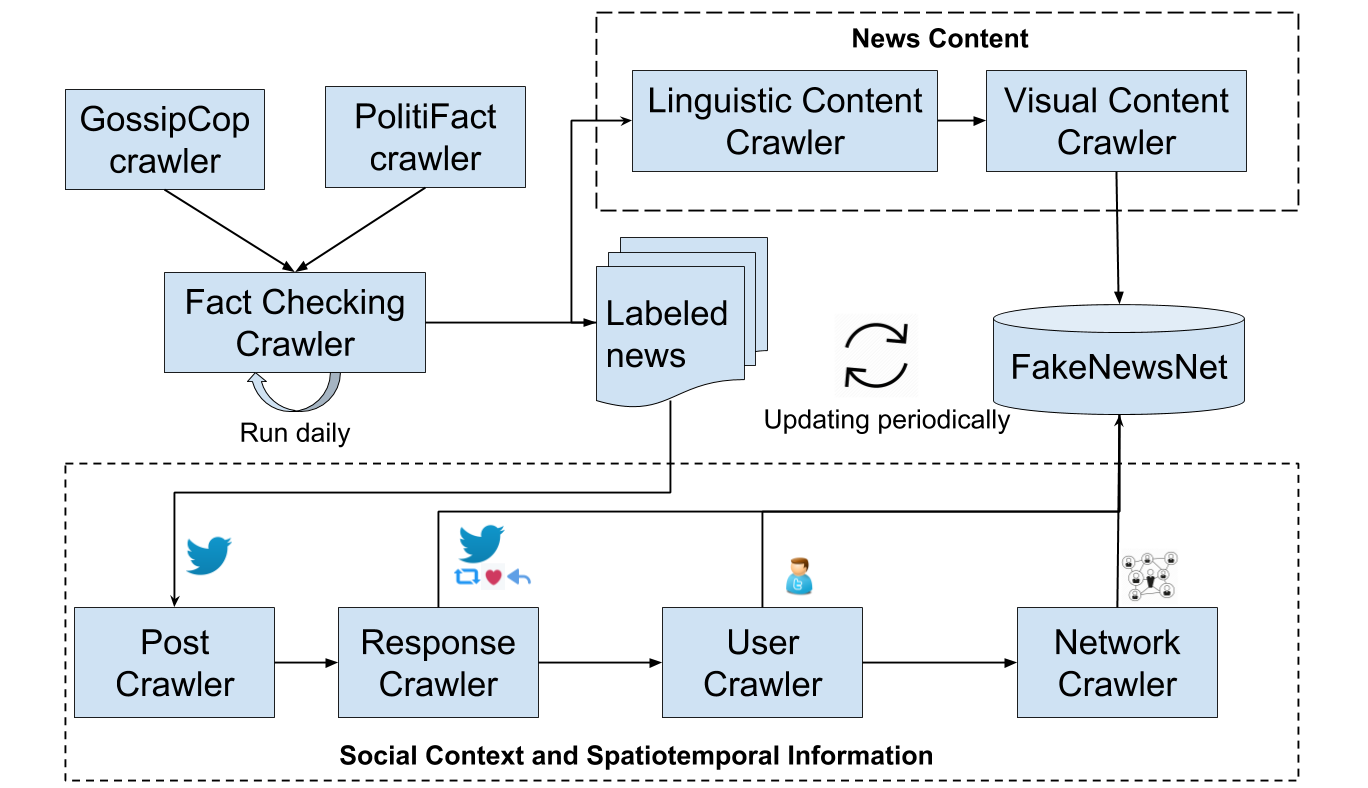}
\caption{The flowchart of dataset integration process for FakeNewsNet. It mainly describes the collection of news content, social context and spatiotemporal information. }
\label{fig:crawler_workflow}}
\end{figure}

In this section, we introduce the dataset integration  process for the FakeNewsNet repository. We demonstrate (see Figure~\ref{fig:crawler_workflow}) how we can collect news contents with reliable ground truth labels, how we obtain additional social context and spatialtemporal information. 
% {\color{blue} shall we use ``collection'' instead of ``construction'', check with Dr. Liu}
% The flowchart of the data collection process is shown in Figure~\ref{fig:crawler_workflow}, which mainly consists of the collection of news contents, social context, and spatiotemporal information. 

% {\color{blue} use the names in table 1 to make connections, figure out the relations among them}

\textbf{News Content}: To collect reliable ground truth labels for fake news, we utilize fact-checking websites to obtain news contents for fake news and true news such as \textit{PolitiFact}~\footnote{https://www.politifact.com/} and \textit{GossipCop}~\footnote{https://www.gossipcop.com/}.
% Ground truth
% \textit{PolitiFact Crawler}
In PolitiFact, journalists and domain experts review the political news and provide fact-checking evaluation results to claim news articles as fake~\footnote{available at https://www.politifact.com/subjects/fake-news/} or real~\footnote{available at https://www.politifact.com/truth-o-meter/rulings/true/}.  We utilize these claims as ground truths for fake and real news pieces.
% News content
In PolitiFact's fact-checking evaluation result, the source URLs of the web page that published the news articles are provided, which can be used to fetch the news contents related to the news articles. In some cases, the web pages of source news articles are removed and are no longer available. To tackle this problem, we i) check if the removed page was archived and automatically retrieve content at the Wayback Machine~\footnote{https://archive.org/web/}; and ii) make use of Google web search in automated fashion to identify news article that is most related to the actual news. 
% PolitiFact is a website operated by Tampa Bay Times, where reporters and editors from the media fact check the political news articles.
%PolitiFact publishes the original statement of news articles, their complete fact-check evaluation results in their website.  
% \textit{GossipCop Crawler}
GossipCop is a website for fact-checking entertainment stories aggregated from various media outlets. GossipCop provides rating scores on the scale of 0 to 10 to classify a news story as the degree from fake to real.  From our observation, almost 90\% of the stories from GossipCop have scores less than 5, which is mainly because the purpose of GossipCop is to showcase more fake stories. In order to collect true entertainment news pieces, we crawl the news articles from E! Online\footnote{https://www.eonline.com/}, which is a well-known trusted media website for publishing entertainment news pieces. We consider all the articles from E! Online as real news sources. We collect all the news stories from GossipCop with rating scores less than 5 as the fake news stories.
% News content

Since GossipCop does not explicitly provide the URL of the source news article, so similarly we search the news headline in Google or the Wayback Machine archive to obtain the news source information. The headline of the GossipCop story articles are generally written to reflect the actual fact and may not be used directly. For example, one of the headlines, ``Jennifer Aniston NOT Wearing Brad Pitts Engagement Ring, Despite Report'' mentions the actual fact instead of the original news articles title. We utilize some heuristics to extract proper headlines such as i) using the text in quoted string; ii) removing negative sentiment words. For example, some headlines include quoted string which are exact text from the original news source. In this case, we extract the named entities through Stanford's CoreNLP tool \cite{manning2014stanford} from the headline and quoted strings from the headline to form the search query. For example, in the headline Jennifer Aniston, Brad Pitt NOT ``Just Married'' Despite Report, we extract named entities including Jennifer Aniston, Brad Pitt and quoted strings including Just Married and form the search query as ``Jennifer Aniston Brad Pitt Just Married'' because the quoted text in addition with named entites mostly provides the context of the original news. As another example, the headline are written in the negative sense to correct the false information, e.g., ``Jennifer Aniston NOT Wearing Brad Pitts Engagement Ring, Despite Report''. So we remove negative sentiment words retrieved from SentiWordNet\cite{baccianella2010sentiwordnet} and some hand-picked words from the headline to form the search query, e.g., ``Jennifer Aniston Wearing Brad Pitts Engagement Ring''. 

\begin{table*}[!htbp]
\small
\begin{center}
\caption{Statistics of the FakeNewsNet repository}\label{tab:dataset_stats}
\begin{tabular}{| l | l | l | c | c | c | c |}
\Xhline{3\arrayrulewidth}
\multirow{2}{*} & \multirow{2}{*} {\textbf{Category}}  &\multirow{2}{*} {\textbf{Features}} & \multicolumn{2}{ c |}{\textbf{PolitiFact}} & \multicolumn{2}{ c |}{\textbf{GossipCop}}   \\ 
\cline{4-7}
& & & \textbf{Fake} & \textbf{Real} & \textbf{Fake} & \textbf{Real}  \\
\Xhline{2\arrayrulewidth}

\multirow{ 3}{*}{\textbf{\makecell{News \\ Content}}} & \multirow{ 2}{*}{\textit{Linguistic}} & \# News articles & 432&  624& 5,323 &  16,817\\ 
% \cline{3-7}

& & \# News articles with text & 420  & 528 & 4,947 &16,694\\ \cline{2-7}
& \textit{Visual} & \# News articles with images & 336  & 447 & 1,650 & 16,767\\ \Xhline{3\arrayrulewidth}

\multirow{12}{*}{\textbf{\makecell{Social \\ Context}}}& \multirow{4}{*}{\textit{User}} & \# Users posting tweets & 95,553 & 249,887 & 265,155 & 80,137\\ %\cline{3-7}
& & \# Users involved in likes &113,473  & 401,363 & 348,852 &145,078\\ %\cline{3-7}
& & \# Users involved in retweets & 106,195 & 346,459& 239,483 & 118,894\\ %\cline{3-7}
& & \# Users involved in replies & 40,585 & 18,6675 & 106,325& 50,799\\ \cline{2-7}

& \textit{Post} & \# Tweets posting news& 164,892 &399,237&  519,581& 876,967\\ 
\cline{2-7}
% & \# News articles with tweets &342   &  314& 4,004 & 2,902\\
% & & News articles with both tweets and text content &286  & 202 &  3,674& 2,895\\ \cline{2-7}

% &\multirow{ 9}{*}{Second order} & News articles with tweets containing at least 1 reply & 236 &  180& 1,657 & 752\\ \cline{3-7}
% & & News articles with tweets containing at least 1 like &283  & 219  & 2,060 &845 \\ \cline{3-7}
% & & News articles with tweets containing at least 1 retweet & 282 &  242 &  2,172 & 1,254\\ \cline{3-7}
&\multirow{ 3}{*}{\textit{Response}}  & \# Tweets with replies & 11,975 &41,852  & 39,717&11,912\\ %\cline{3-7}
& & \# Tweets with likes & 31692 & 93,839&96,906& 41,889\\ %\cline{3-7}
& & \# Tweets with retweets & 23,489 &67,035 & 56,552& 24,955\\ \cline{2-7}

& \multirow{4}{*}{\textit{Network}} & \# Followers &  405,509,460 &1,012,218,640& 630,231,413& 293,001,487\\ %\cline{3-7}
& & \# Followees &449,463,557& 1,071,492,603 &619,207,586 & 308,428,225\\ %\cline{3-7}
& & Average \# followers & 1299.98 & 982.67& 1020.99& 933.64\\ %\cline{3-7}
& & Average \# followees & 1440.89 & 1040.21 & 1003.14&  982.80\\ \Xhline{3\arrayrulewidth}

\multirow{5}{*}{\textbf{\makecell{Spatiotemporal \\ Information}}}&\multirow{2}{*}{\textit{Spatial}} & \# User profiles with locations  & 217,379 & 719,331& 429,547 & 220,264\\
&&\# Tweets with locations  & 3,337 & 12,692 & 12,286 & 2,451\\
\cline{2-7}
& \multirow{2}{*}{\textit{Temporal}} & \# Timestamps for news pieces & 296 & 167 & 3,558 & 9,119 \\ 
% & & \# Timestamps for posts &116,005 & 261,262  &  487,327 & 154,383\\ 
& & \# Timestamps for response &171,301 & 669,641& 381,600 & 200,531 \\ 
\Xhline{2\arrayrulewidth}
\end{tabular}
\end{center}
\end{table*}

\textbf{Social Context}: The user engagements related to the fake and real news pieces from fact-checking websites are collected using search API provided by social media platforms such as the Twitter's Advanced Search API~\footnote{https://twitter.com/search-advanced?lang=en}. The search queries for collecting user engagements are formed from the headlines of news articles, with special characters removed from the search query to filter out the noise. After we obtain the social media posts that directly spread news pieces, we further fetch the user \textit{response} towards these posts such as replies, likes, and reposts. In addition, when we obtain all the users engaging in news dissemination process, we collect all the metadata for user profiles, user posts, and the social network information.

\textbf{Spatiotemporal Information}: The spatiotemporal information includes spatial and temporal information. For spatial information, we obtain the locations explicitly provided in user profiles. The temporal information indicates that we record the timestamps of user engagements, which can be used to study how fake news pieces propagate on social media, and how the topics of fake news are changing over time. Since fact-checking websites periodically update newly coming news articles, so we dynamically collect these newly added news pieces and update the FakeNewsNet repository as well. In addition, we keep collecting the user engagements for all the news pieces periodically in the FakeNewsNet repository such as the recent social media posts, and second order user behaviors such as replies, likes, and retweets. For example, we run the news content crawler and update Tweet collector per day. The spatiotemporal information provides useful and comprehensive information for studying fake news problem from a temporal perspective.

\section{Data Analysis }\label{sec:analysis}

% {\color{blue} Purpose of data analysis: (1) show that the datasets makes sense. For example, follows power law distribution, can give reasonable fake news detection results; (2) it contains abundant information for conducting various fake news detection tasks. Try to make more connections to (1)}
% {\color{blue} multi-dimension or multi-dimensional?}
FakeNewsNet has multi-dimensional information related to news content, social context, and spatiotemporal information. In this section, we first provide some preliminary quantitative analysis to illustrate the features of FakeNewsNet. We then perform fake news detection using several state-of-the-art models to evaluate the quality of the FakeNewsNet repository. The detailed statistics of FakeNewsNet repository is illustrated in Table~\ref{tab:dataset_stats}.

\subsection{Assessing News Content}
\begin{figure}[!h]
\centering
\subfigure[PolitiFact Fake News]{
 {\includegraphics[scale=0.29]{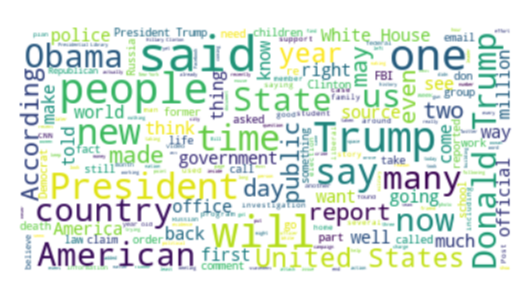}
\label{fig:politifact_fake_word_cloud}
}
}
\subfigure[PolitiFact Real News]{
 {\includegraphics[scale=0.29]{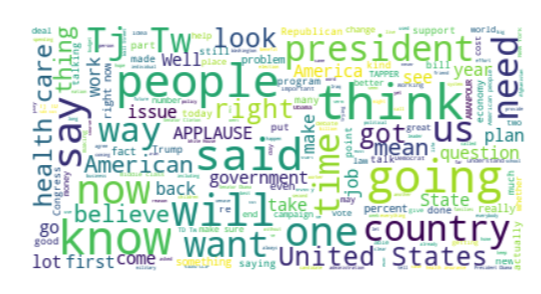}
 \label{fig:politifact_real_word_cloud}
}
}
\subfigure[GossipCop Fake News]{
 {\includegraphics[scale=0.285]{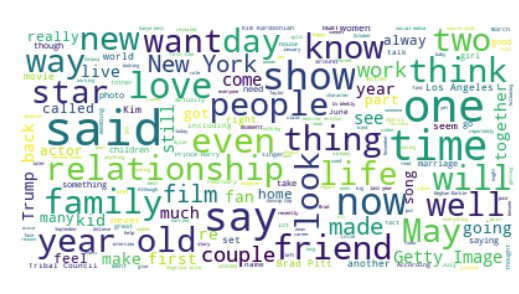}
 \label{fig:gossipcop_fake_word_cloud}
}
}
\subfigure[GossipCop Real News]{
 {\includegraphics[scale=0.285]{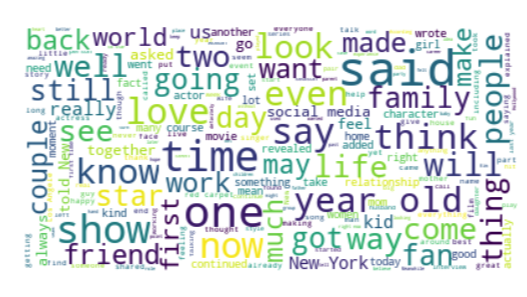}
 \label{fig:gossipcop_real_word_cloud}
}
}
\vskip -1em
\caption{The word cloud of new body text for fake and real news on PolitiFact and GossipCop.}
\label{fig:word_cloud}
\vspace{-0.1cm}
\end{figure}
\begin{figure*}[!h]
\centering
\subfigure[PolitiFact dataset]{
 {\includegraphics[scale=0.23]{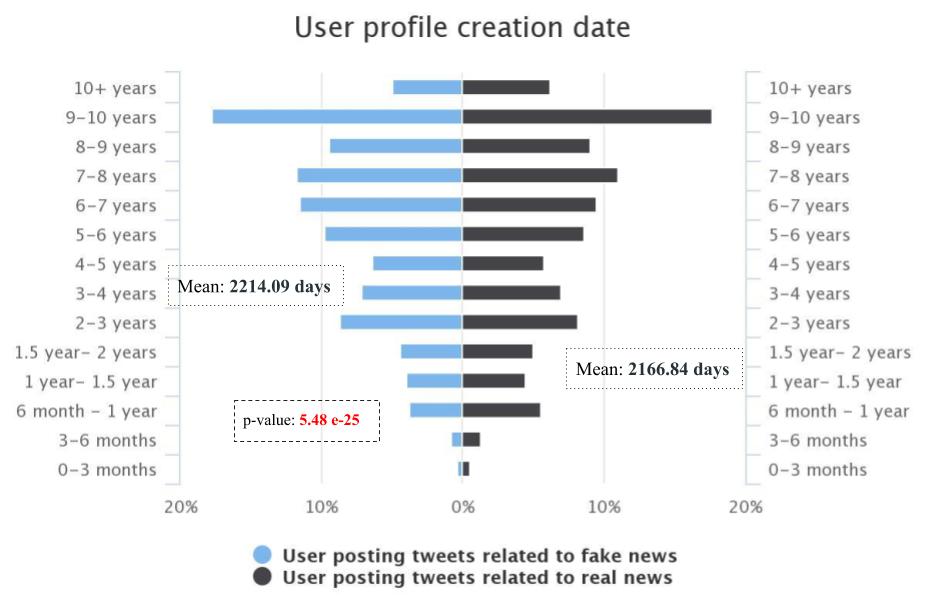}}
}
\subfigure[GossipCop dataset]{
 {\includegraphics[scale=0.23]{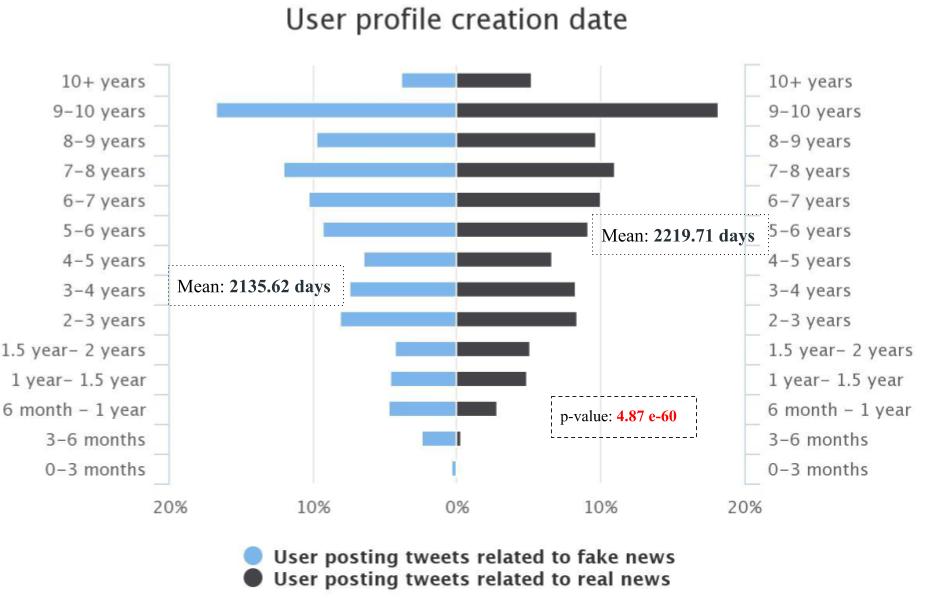}}
}
\vskip -0.5em
\caption{The distribution of user profile creation dates on PolitiFact and GossipCop datasets}
\label{fig:user_profile_creation}
\vspace{-0.1cm}
\end{figure*}
Since fake news attempts to spread false claims in news content, the most straightforward means of detecting it is to find clues in a news article to detect fake news. 
%News content features describe the meta information related to a piece of news. A list of representative news content attributes include publishers, headlines, body texts, and videos.  
First, we analyze the topic distribution of fake and real news articles. From figures \ref{fig:politifact_fake_word_cloud} and \ref{fig:politifact_real_word_cloud}, we can observe that the fake and real news of the PolitiFact dataset is mostly related to the political campaign. In case of GossipCop dataset from figures \ref{fig:gossipcop_fake_word_cloud} and \ref{fig:gossipcop_real_word_cloud}, we observe that the fake and real news are mostly related to gossip about the relationship among celebrities. In addition, we can see the topics for fake news and real news are slightly different in general. However, for specific news, it is difficult to only use topics in the content to detect fake news~\cite{shu2017fake}, which necessitates the need to utilize other auxiliary information such as social context.

We also explore the distribution of publishers who publish fake news on both datasets. We find out that there are in total 301 publishers publishing 432 fake news pieces, among which 191 of all publishers only publish 1 piece of fake news, and 40 publishers publish at least 2 pieces of fake news such as theglobalheadlines.net and worldnewsdailyreport.com. For Gossipcop, there are in total 209 publishers publishing 6,048 fake news pieces, among which 114 of all publishers only publish 1 piece of fake news, and 95 publishers publish at least 2 pieces of fake news such as hollywoodlife.com and celebrityinsider.org. The reason may be that these fact-checking websites try to identify those check-worthy breaking news events regardless of the publishers, and fake news publishers can be shut down after they were reported to publish fake news pieces.

% \begin{table*}[!h]
% \begin{center}
% \caption{The statistics of the social network of the datasets}\label{tab:social_net_stats}
% \begin{tabular}{| c | c | c | c | c |}
% \hline
% \multirow{2}{*}{\diagbox{\textbf{Features}}{\textbf{Dataset}}} & \multicolumn{2}{ c |}{\textbf{PolitiFact}} & \multicolumn{2}{ c |}{\textbf{GossipCop}}   \\ 
% \cline{2-5}
% & \textbf{Fake} & \textbf{Real} & \textbf{Fake} & \textbf{Real}  \\
% \hline
% \# Users &214,484& 694,981 & 697,217 & 69242\\ \hline
% \# Followers & 262,183,502 &711,289,100& 530,571,428 & 73,485,774\\ \hline
% \# Followees &  287,997,894& 742,828,754 &529,517,989 & 74,655,172\\ \hline
% Avg. followers & 1222.392 & 1023.465 & 760.984 &1061.289 \\ \hline
% Avg. followees &  1342.748 & 1068.847 & 759.474 & 1078.177\\ \hline
% \end{tabular}
% \end{center}
% \end{table*}

\subsection{Measuring Social Context}
Social context represents the news proliferation process over time, which provides useful auxiliary information to infer the veracity of news articles.  Generally, there are three major aspects of the social media context that we want to represent: user profiles, user posts, and network structures. Next, we perform an exploratory study of these aspects on FakeNewsNet and introduce the potential usage of these features to help fake news detection.

\subsubsection{User Profiles}
User profiles on social media have been shown to be correlated with fake news detection~\cite{shu2018understanding}. Research has also shown that fake news pieces are likely to be created and spread by non-human accounts, such as social bots or cyborgs~\cite{shu2017fake,shao2017spread}. We will illustrate some user profile features in FakeNewsNet repository.

% \begin{figure}[tbp!]
% \centering
% \includegraphics[scale=0.23]{politifact_user_profile_creation_date.jpg}
% \vskip -1em
% \caption{User profile creation dates on PolitiFact}
% \label{fig:user_profile_creation}
% \end{figure}

\begin{figure}[!h]
\centering
\includegraphics[scale=0.3]{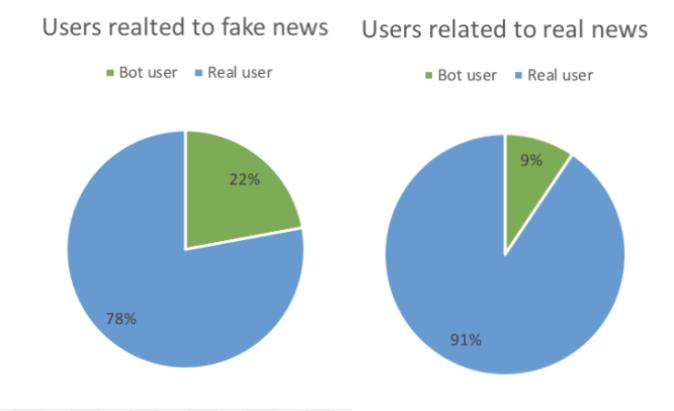}
\vskip -1em
\caption{Comparison of bot scores on users related to fake and real news on PolitiFact dataset.}
\label{fig:bot_analysis}
\end{figure}
First, we explore whether the creation time of user accounts for fake news and true news are different or not. We compute the time range of account register time with the current date and the results are shown in Figure~\ref{fig:user_profile_creation}. We can see that the account creation time distribution of users who posting fake news is significantly different from those who post real news, with the p-value$<0.05$ under statistical t-test. In addition, we notice that it's not necessary that users with an account created long time or shorter time post fake/real news more often. For example, the mean creation time for users posting fake news ($2214.09$) is less than that for real news ($2166.84$) in Politifact; while we see opposite case in Gossipcop dataset.

%users who share real news pieces tend to have longer register time than those who share the fake news in average. For example, around $19\%$ of all users who share real news pieces have registered for 9 to 10 years, while only $15\%$ for those who share fake news.  The reason could be that these newly created accounts are created intentionally to spread fake news such as social bots or sybils~\cite{shu2017fake}.
% {\color{blue} REVISING the claim}
% {\color{blue} but for 4~9 years, the observation is ``users who share real news pieces tend to have longer register time than those who share the fake news'' is not true. The claim is questionable.}

Next, we take a deeper look into the user profiles and assess the social bots effects. We randomly selected 10,000 users who posted fake and real news and performed bot detection using one of the state-of-the-art bot detection algorithm Botometer \cite{davis2016botornot} API\footnote{https://botometer.iuni.iu.edu/}. The Botometer takes a Twitter username as input and utilizes various features extracted from meta-data and output a probability score in $[0,1]$, indicating how likely the user is a social bot. We set the threshold of 0.5 on the bot score returned from the Botometer results to determine bot accounts.  Figure \ref{fig:bot_analysis} shows the ratio of the bot and human users involved in tweets related to fake and real news. We can see that bots are more likely to post tweets related to fake news than real users. For example, almost 22\% of users involved in fake news are bots, while only around $9\%$ of users are predicted as bot users for real news. Similar results were observed with different thresholds on bot scores based on both datasets. This indicates that there are bots in Twitter for spreading fake news, which is consistent with the observation in~\cite{shao2017spread}. In addition, most users that spread fake news (around 78\%) are still more likely to be humans than bots (around 22\%), which is also in consistence with the findings in~\cite{vosoughi2018spread}.

\subsubsection{Post and Response}
People express their emotions or opinions towards fake news through social media posts, such as skeptical opinions, sensational reactions, etc. These features are important signals to study fake news  and disinformation in general~\cite{jin2016news,qazvinian2011rumor}.

\begin{figure}[tbp]
\centering
\subfigure[PolitiFact dataset]{
 {\includegraphics[scale=0.13]{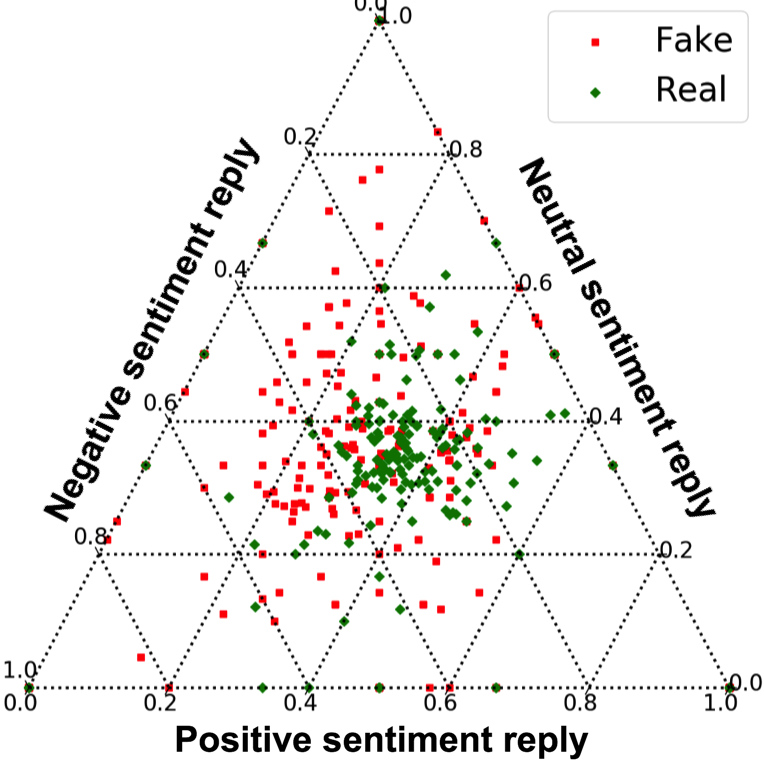}}
 \label{fig:politifact_sentiment}
}
\subfigure[GossipCop dataset]{
 {\includegraphics[scale=0.13]{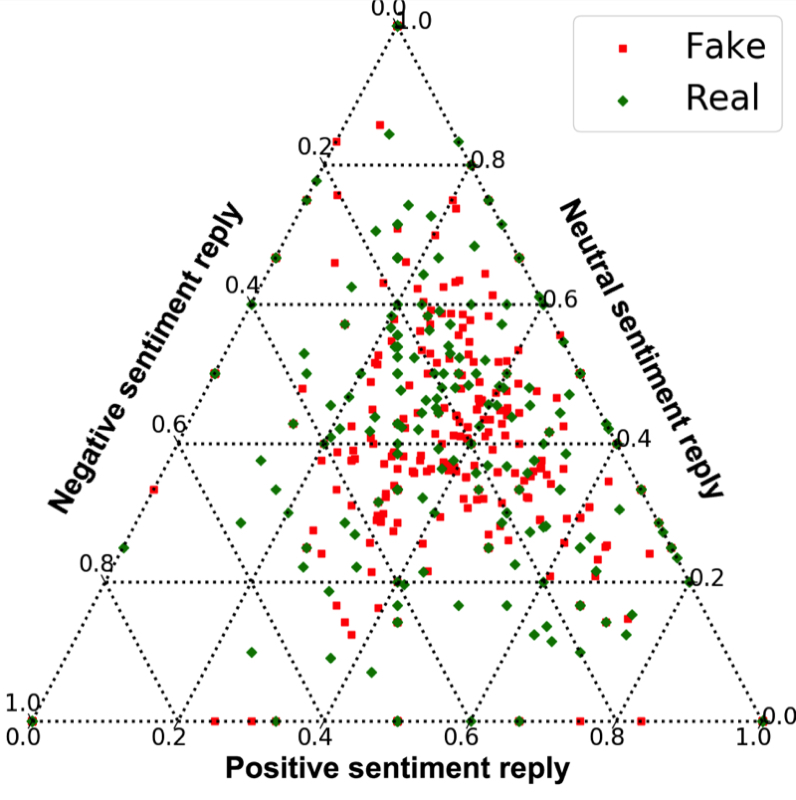}}
 \label{fig:gossipcop_sentiment}
}
\vskip -0.5em
\caption{Ternary plot of the ratio of the positive, neutral and negative sentiment replies for fake and real news.}
\label{fig:user_feature_analysis}
\vspace{-0.1cm}
\end{figure}
We perform sentiment analysis on the replies of user posts that spreading fake news and real news using one of the state-of-the-art unsupervised prediction tool called VADER~\footnote{https://github.com/cjhutto/vaderSentiment}~\cite{gilbert2014vader}. It is a lexicon and rule-based sentiment analysis tool that is specifically attuned to sentiments expressed in social media. Figure \ref{fig:user_feature_analysis} shows the relationship between positive, neutral and negative replies for all news articles. For each news piece, we obtain all the replies for this news piece and predict the sentiment as positive, negative, or neutral. Then we calculate the ratio of positive, negative, and neutral replies for the news. For example, if a news piece has the sentiment distribution of replies as $[0.5,0.5,0.5]$, it occurs in the middle of the very center of the triangle in Figure~\ref{fig:politifact_sentiment}. We can also see that the real news have more number of neutral replies over positive and negative replies whereas fake articles have a bigger ratio of negative sentiments. In case of sentiment of the replies of the Gossipcop dataset shown in Figure \ref{fig:gossipcop_sentiment}, we cannot observe any significant differences between fake and real news. This could be because of the difficulty in identifying fake and real news related to entertainment by common people. 
% {\color{blue} sample equal number of fake and real news to see if any patterns} 

We analyze the distribution of the likes, retweets, and replies of the tweets, which can help gain insights on user interaction networks related to fake and real news. Social science studies have theorized the relationship between user behaviors and their perceived beliefs on the information on social media~\cite{kim2017says}. For example, the behaviors of likes and retweets are more emotional while replies are more rational.
% \begin{figure}[!htbp]
% \centering
% \subfigure[Tweets related to fake news in PolitiFact dataset]{
%  {\includegraphics[scale=0.21]{politifact_fake_tweet_post_heatmap.png}}
% }
% \subfigure[Tweets related to real news in PolitiFact dataset]{
%  {\includegraphics[scale=0.21]{politifact_real_tweet_post_heatmap.png}}
% }

% % \subfigure[Tweets related to fake news in Gossipcop dataset]{
% %  {\includegraphics[scale=0.23]{gossipcop_fake_tweet_post_heatmap.png}}
% % }
% % \subfigure[Tweets related to real news in Gossipcop dataset]{
% %  {\includegraphics[scale=0.23]{gossipcop_real_tweet_post_heatmap.png}}
% % }
% \caption{The heatmap of the day of week vs hour of tweets posted related to fake and real news}
% \label{fig:timeline_heatmap }
% \vspace{-0.1cm}
% \end{figure}

\begin{figure}[!tbp]
\centering
\subfigure[PolitiFact dataset]{
 {\includegraphics[scale=0.12]{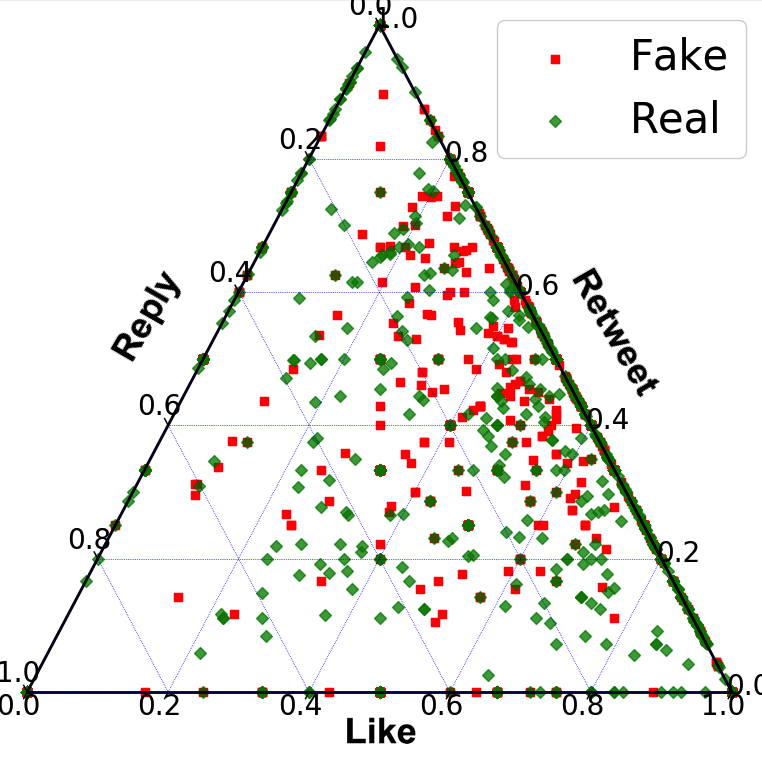}}
}
\subfigure[GossipCop dataset]{
 {\includegraphics[scale=0.12]{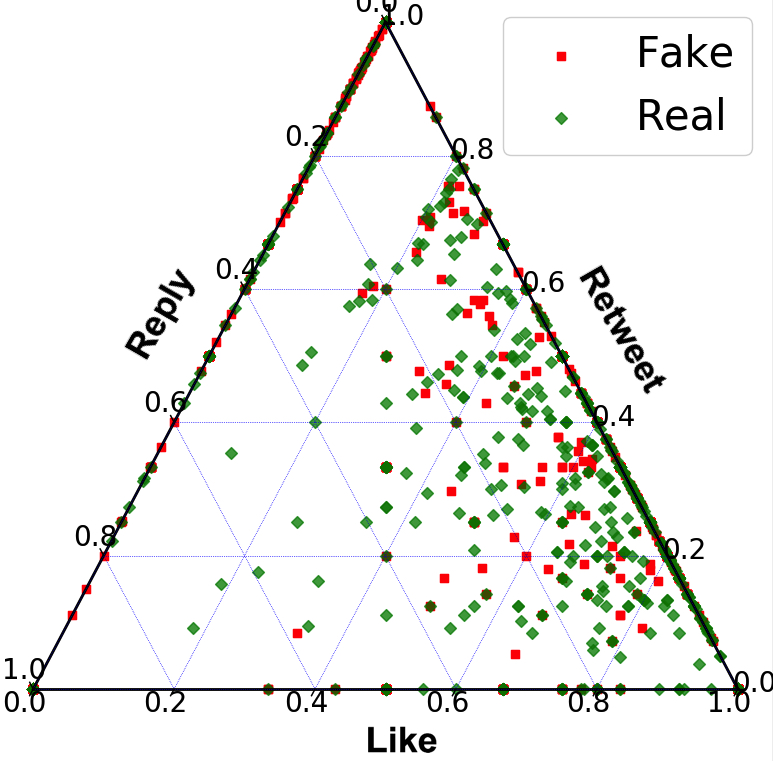}}
}
\vskip -0.5em
\caption{Ternary plot of the ratio of likes, retweet and reply of tweets related to fake and real news}
\label{fig:ternary_plot}
\vspace{-0.1cm}
\end{figure}

We plot the ternary triangles which illustrate the ratio of replies, retweets, and likes from the second order engagements towards the posts that spread fake news or real news pieces. From Figure \ref{fig:ternary_plot}, we observe that the: i) fake news pieces tend to have fewer replies and more retweets; ii) Real news pieces have more ratio of likes than fake news pieces, which may indicate that users are more likely to agree on real news. The differences in the distribution of user behaviors between fake news and real news have potentials to study users' beliefs characteristics. FakeNewsNet provides real-world datasets to understand the social factors of user engagements and underly social science as well.

\subsubsection{Network}
Users tend to form different networks on social media in terms of interests, topics, and relations, which serve as the fundamental paths for information diffusion~\cite{shu2017fake}.  Fake news dissemination processes tend to form an echo chamber cycle, highlighting the value of extracting network-based features to represent these types of network patterns for fake news detection~\cite{del2016echo}.

\begin{figure}[!h]
\centering
\subfigure[Follower count of users in PolitiFact dataset]{
 {\includegraphics[scale=0.17]{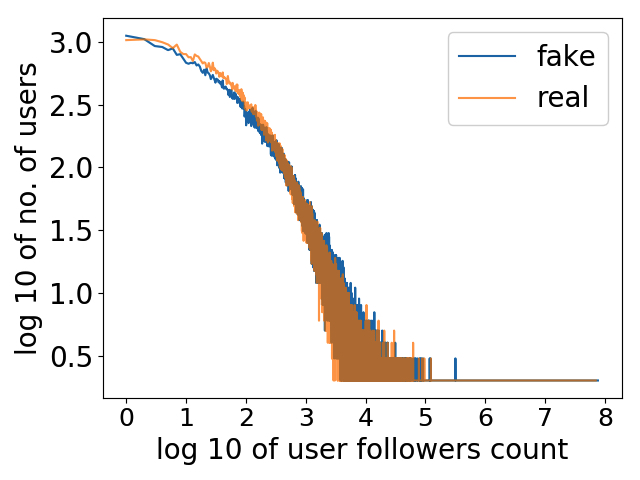}}
 \label{fig:politifact_follower}
	}
\subfigure[Followee count of users in PolitiFact dataset]{
 {\includegraphics[scale=0.17]{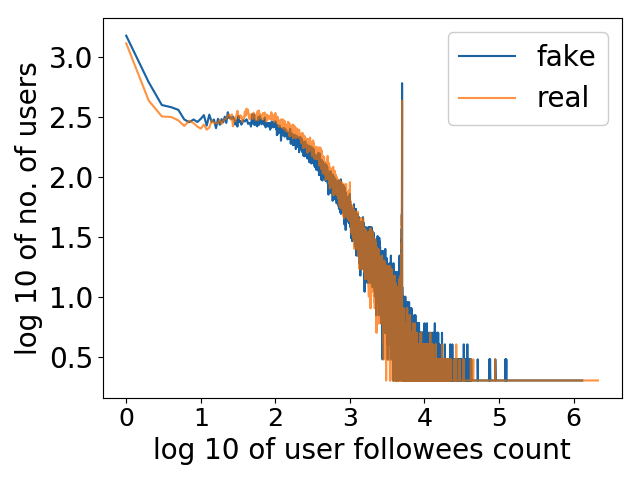}}
 \label{fig:user_followee}
}
\subfigure[Follower count of users in Gossipcop dataset]{
 {\includegraphics[scale=0.17]{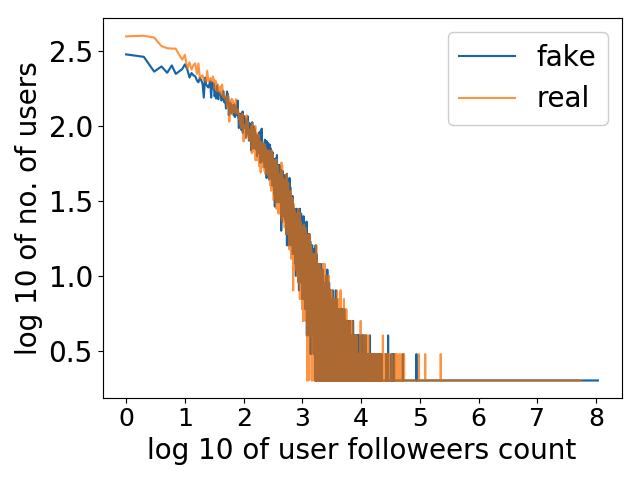}}
 \label{fig:gossip_follower}
	}
\subfigure[Followee count of users in PolitiFact dataset]{
 {\includegraphics[scale=0.17]{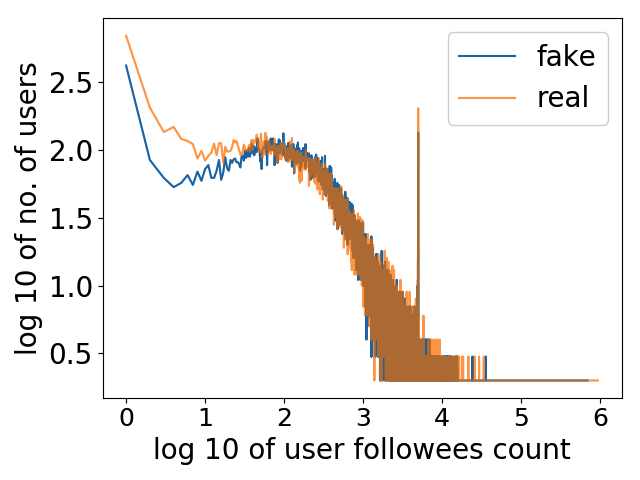}}
 \label{fig:gossip_followee}
}
\vskip -0.5em
\caption{The distribution of the count of followers and followees related to fake and real news}
\label{fig:social_network}
\vspace{-0.1cm}
\end{figure}

We look at the social network statistics of all the users that spread fake news or real news. The social network features such as  followers count and followee count can be used to estimate the scope of how the fake news can spread in social media. We plot the distribution of follower count and followee count of users in Figure~\ref{fig:social_network}. We can see that: i) the follower and followee count of the users generally follows power law distribution, which is commonly observed in social network structures; ii) there is a spike in the followee count distribution of both users and this is because of the restriction imposed by Twitter\footnote{https://help.twitter.com/en/using-twitter/twitter-follow-limit} on users to have at most 5000 followees when the number of following is less than 5000.

\subsection{Characterizing Spatiotemporal Information}

\begin{figure}[!htbp]
\centering
\subfigure[Temporal user engagements of fake news]{
 {\includegraphics[scale=0.25]{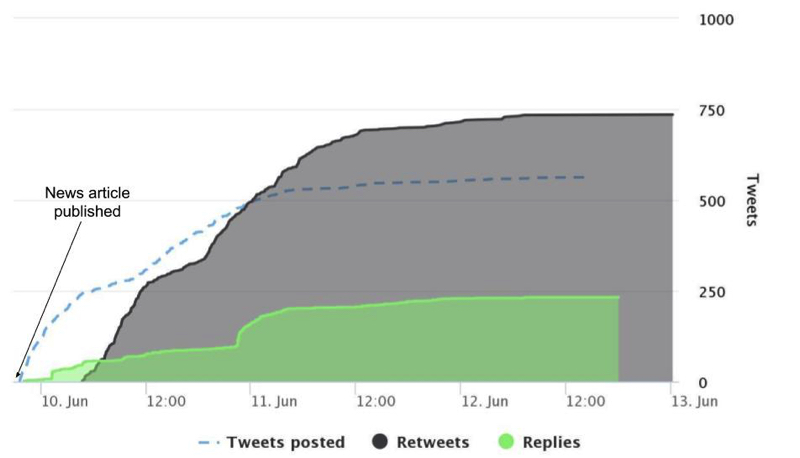}}
 \label{fig:fake_dynamic_info}
}
\subfigure[Temporal user engagements of real news]{
 {\includegraphics[scale=0.25]{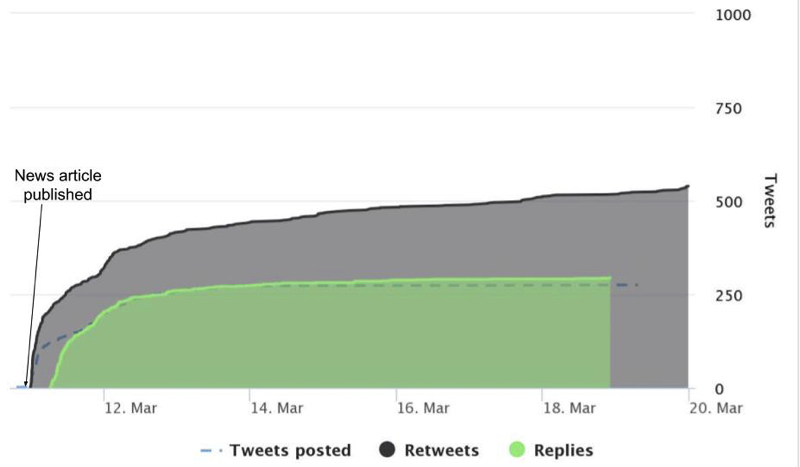}
 \label{fig:real_dynamic_info}}
}
\vskip -0.5em
\caption{The comparison of temporal user engagements of fake and real news}\label{fig:dynamic_information}\vspace{-0.1cm}
\end{figure}

Recent research has shown users' temporal responses can be modeled using deep neural networks to help detection fake news~\cite{ruchansky2017csi}, and deep generative models can generate synthetic user engagements to help early fake news detection~\cite{liu2018early}. The spatiotemporal information in FakeNewsNet depicts the temporal user engagements for news articles, which provides the necessary information to further study the utility of using spatiotemporal information to detect fake news.

% First, we look at whether there are differences on the temporal behaviors of users for spreading fake news and real news. We illustrate the relationship between the posting time and day of the week as in Figure \ref{fig:timeline_heatmap}. We can see that the time period at which the tweets related to fake news and real news posted are different. For example, in case of fake news, tweets are posted night even at odd hours between 1 am to 5 am where people are generally inactive and the density of tweets are almost the same as peak hours. This could generally because of bot accounts programmed running throughout the day. 

% \begin{figure}[!h]
% \centering
% \includegraphics[scale=0.25]{tweets_timeline.png}
% \caption{Timeline of the tweets in the dataset}
% \label{fig:tweet_timeline}
% \end{figure}

First, we investigate if the temporal user engagements such as posts, replies, retweets, are different for fake news and real news with similar topics, e.g., fake news ``\textit{TRUMP APPROVAL RATING Better than Obama and Reagan at Same Point in their Presidencies}'' from June 9, 2018 to 13 June, 2018 and real news ``\textit{President Trump in Moon Township Pennsylvania}'' from March 10, 2018 to 20 March, 2018. As shown in Figure~\ref{fig:dynamic_information}, we can observe that: i) for fake news, there is a sudden increase in the number of retweets and it does remain constant beyond a short time whereas, in the case of real news, there is a steady increase in the number of retweets; ii) Fake news pieces tend to receive fewer replies than real news. We have similar observations in Table~\ref{tab:dataset_stats}, and replies count for $5.76\%$ among all Tweets for fake news, and $7.93\%$ for real news. The differences of diffusion patterns for temporal user engagements have the potential to determine the threshold time for early fake news detection. For example, if we can predict the sudden increase of user engagements, we should use the user engagements before the time point and detect fake news accurately to limit the affect size of fake news spreading~\cite{fakebookchapter}.

\begin{figure}[!tbp]
\centering
\subfigure[Spatial distribution of users posting fake news]{
 {\includegraphics[scale=0.3]{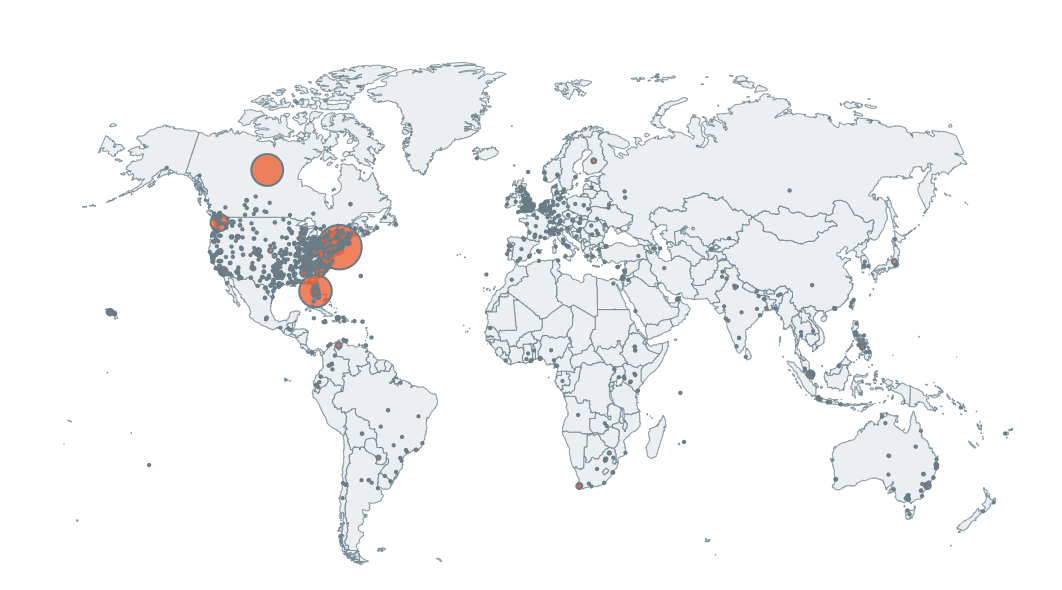}}
 \label{fig:fake_dynamic_info}
}
\subfigure[Spatial distribution of users posting real news]{
 {\includegraphics[scale=0.3]{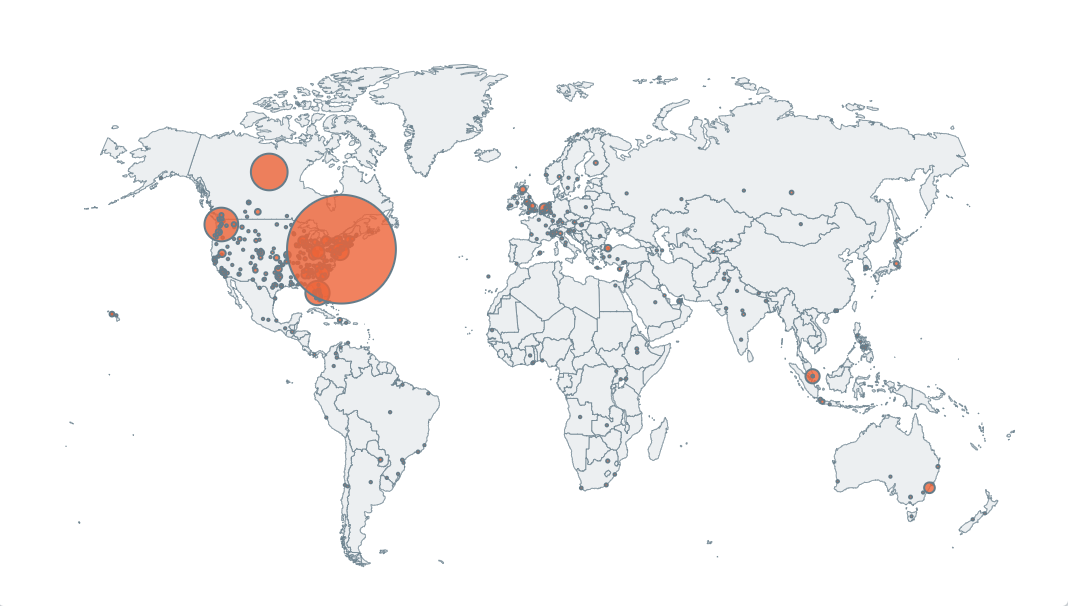}
 \label{fig:real_dynamic_info}}
}
\vskip -0.5em
\caption{Spatial distribution of users posting tweets related to fake and real news.}\label{fig:spatial_information}\vspace{-0.1cm}
\end{figure}

Next, we demonstrate the geo-location distribution of users engaging in fake and real news (See Figure~\ref{fig:spatial_information} for Politifact dataset). We show the locations explicitly provided by users in their profiles, and we can see that users in the PolitiFact dataset who posting fake news have a different distribution than those posting real news. Since it is usually sparse of locations provided by users explicitly, we can further consider the location information attached with Tweets, and even utilize existing approaches for inferring the locations~\cite{zubiaga2017towards}. It would be interesting to explore how users are geo-located distributes using FakeNewsNet repository from different perspectives.

% {\color{blue} describe the figure 9, is it for PolitiFact?} 
% Tweets location prediction~\cite{zubiaga2017towards}

\subsection{Fake News Detection Performance}
In this subsection, we utilize the PolitiFact and GossipCop datasets from FakeNewsNet repository to perform fake news detection. We use $80\%$ of data for training and $20\%$ for testing. For evaluation metrics, we use accuracy, precision, recall and F1 score. We deployed several state-of-the-art baselines for fake news detection,
%In order to evaluate different dimensions of the dataset, we choose baselines that 1) only consider news content; 2) only consider social context; and 3) both news content and social context. 
%Since we evaluate both dimension, we considered samples that has both news content and user engagements with atleast 1 reply for all the experiments performed.

\begin{itemize}
\item \textbf{News content:}
%News content includes the articles of the source web pages of fake and real news. We utilize the raw text features of the news articles and represent them as one-hot vectors. %Only the first 500 words of the news articles are considered for the classification. 
% {\color{blue} What are the extracted features? How to extract the feature from news content}
To evaluate the news contents, the text contents from source news articles are represented as a one-hot encoded vector and then we apply standard machine learning models including support vector machines (SVM), logistic regression (LR), Naive Bayes (NB), and CNN. For SVM, LR, and NB, we used the default settings provided in the scikit-learn and do not tune parameters. For CNN we use the standard implementation with default setting~\footnote{https://github.com/dennybritz/cnn-text-classification-tf}. We also evaluate the classification of news articles using Social article fusion (SAF /S) \cite{shu2018fakenewstracker} model that utilizes auto-encoder for learning features from news articles to classify new articles as fake or real.

\item \textbf{Social context:}
In order to evaluate the social context, we utilize the variant of SAF model~\cite{shu2018fakenewstracker}, i.e., SAF /A, which utilize the temporal pattern of the user engagements to detect fake news. 
%The models uses the text content and latent user representation of the replies.

\item \textbf{News content and social context:}
Social Article Fusion(SAF) model that combines SAF /S and SAF /A is used. This model uses autoencoder with LSTM cells of 2 layers for encoder as well as decoder and also temporal pattern of the user engagements are also captured using another network of LSTM cells with 2 layers.  

\end{itemize}
The experimental results are shown in Table~\ref{tab:res}. We can see that: i) For news content-based methods, SAF /S perform better in terms of accuracy, recall, and F1 score while logistic regression has better precision than others. SAF /A provides a similar result around 66.7\% accuracy as SAF /S but has higher precision. The compared baselines models provide reasonably good performance results for the fake news detection where accuracy is mostly around 65\% on PolitiFact; ii) we observe that SAF relatively achieves better accuracy than both SAF /S and SAF /A for both dataset. For example, SAF has around 5.65\% and 3.68\% performance improvement than SAF /S and SAF /A on PolitiFact. This indicates that user engagements can help fake news detection in addition to news articles on PolitiFact dataset.

In summary, FakeNewsNet provides multiple dimensions of information that has the potential to benefit researchers to develop novel algorithms for fake news detection.

% \begin{table*}[!h]
% \label{tab:res}
% \begin{center}
% \caption{Fake news detection performance on FakeNewsNet}
% \label{tab:baseline_perf}
% \begin{tabular}{| p{40mm} | c | c | c |c | c | c | c |c |}
% \hline
% \textbf{Model} & \multicolumn{4}{  c |}{\textbf{PolitiFact}} & \multicolumn{4}{  c |}{\textbf{GossipCop}}   \\ 
% \cline{2-9}
% & \textbf{Accuracy} & \textbf{Precision} & \textbf{Recall} & \textbf{F1} &\textbf{Accuracy} & \textbf{Precision} & \textbf{Recall} & \textbf{F1}  \\
% \hline

% SVM & 0.580 & 0.611 & 0.717 &0.659  & 0.593 &0.561 & 0.892&  0.689\\ \hline
% Logistic regression & 0.642 & 0.757 &0.543 &0.633 & 0.793&  0.843& 0.724 & 0.779\\ \hline
% Naive Bayes &  0.617  &  0.674 & 0.630 &0.651 &0.861 & 0.885&0.832 &0.858  \\ \hline
% CNN &  0.629&0.807 & 0.456&  0.583  & 0.777& 0.925 & 0.637& 0.755  \\ \hline
% Social Article Fusion /S &0.654  & 0.600 & 0.789&0.681 & 0.651 & 0.637  & 0.537   &0.583   \\ \hline
% Social Article Fusion /A &0.667  &0.667& 0.579 & 0.619 &  0.772 &  0.756& 0.739 &  0.747\\ \hline
% Social Article Fusion & 0.691&0.638  &0.789  & 0.706 &  0.786&  0.808& 0.694 &0.747 \\ \hline
% \end{tabular}
% \end{center}
% \end{table*}

\begin{table*}[!h]
\label{tab:res}
\begin{center}
\caption{Fake news detection performance on FakeNewsNet}
\label{tab:baseline_perf}
\begin{tabular}{| p{40mm} | c | c | c |c | c | c | c |c |}
\hline
\textbf{Model} & \multicolumn{4}{  c |}{\textbf{PolitiFact}} & \multicolumn{4}{  c |}{\textbf{GossipCop}}   \\ 
\cline{2-9}
& \textbf{Accuracy} & \textbf{Precision} & \textbf{Recall} & \textbf{F1} &\textbf{Accuracy} & \textbf{Precision} & \textbf{Recall} & \textbf{F1}  \\
\hline

SVM & 0.580 & 0.611 & 0.717 &0.659  & 0.497 &0.511&   0.713&  0.595\\ \hline
Logistic regression & 0.642 & 0.757 &0.543 &0.633 & 0.648 & 0.675 &  0.619 & 0.646\\ \hline
Naive Bayes &  0.617  &  0.674 & 0.630 &0.651 &0.624 & 0.631 & 0.669 & 0.649 \\ \hline
CNN &  0.629&0.807 & 0.456&  0.583  & 0.723 & 0.751&  0.701 &  0.725  \\ \hline
Social Article Fusion /S &0.654  & 0.600 & 0.789&0.681 & 0.689 & 0.671 & 0.738&  0.703 \\ \hline
Social Article Fusion /A &0.667  &0.667& 0.579 & 0.619 & 0.635 & 0.589&  0.882&   0.706\\ \hline
Social Article Fusion & 0.691&0.638  &0.789  & 0.706 & 0.689 & 0.656&  0.792&   0.717 \\ \hline
\end{tabular}
\end{center}
\end{table*}
% {\color{blue} can you change the numbers of Gossip Performance since we have the gossipcop data and partial of the user engagements}

\section{Data Structure}\label{sec:structure}
In this section, we describe in details of the structure of FakeNewsNet. We will introduce the data format and provided API interfaces that allows for efficient slicing of the data.

% We provide Python scripts which will perform all of the necessary content acquisition, API requests, creation of directory hierarchies on disk, and writing of the data. the user will be able to obtain the entire dataset at their disposal, along with a custom-made API that allows for efficient slicing of the data.

\subsection{Data Format}
Each directory will possess the associated autogenerated news ID as its name and contain the following structure: news content.json file, tweets folder, retweets folder. Finally, user\_profiles folder and user\_timeline\_tweets folder contains the user profile information about all the users involved in tweet provided in the dataset.

\begin{itemize}
\item \textit{news content.json}
includes all the meta information of the news articles collected using the provided news source URLs. This is a JSON object with attributes including:

\begin{itemize}
\item \textit{text} is the text of the body of the news article.
\item \textit{images} is a list of the URLs of all the images in the news article web page.
\item \textit{publish date} indicate the date that news article is published.
\end{itemize}

\item \textit{tweets folder} contains the metadata of the list of tweets associated with the news article collected as separate files for each tweet. Each file in this folder contains the tweet objects returned by Twitter API.

\item \textit{retweets folder}
includes a list of files containing the retweets of tweets associated with the news article. Each file is named as $<$tweet id$>$.json and have a list of retweet objects associated with a particular tweet collected using Twitter API.

% includes all the metadata of the retweets collected using Twitter API. This is a JSON object with tweet id as key and the value is a list of retweet objects collected using Twitter API.

\item \textit{user\_profiles folder}
includes files containing all the metadata of the users in the dataset. Each file is this directory is a JSON object collected from Twitter API containing information about the user including profile creation time, geolocation of the user, profile image URL, followers count, followees count, number of tweets posted and number of tweets favorited.

\item \textit{user\_timeline\_tweets folder} includes JSON files containing the list of at most 200 recent tweets posted by the user. This includes the complete tweet object with all information related to tweet.

% \begin{itemize}
% \item \textit{user\_timeline\_tweets folder} is a JSON object collected from Twitter API containing information about the user including profile creation time, geolocation of the user, profile image URL, followers count, followees count, number of tweets posted and number of tweets favorited.

% \item \textit{recent tweets} is the list of at most 200 recent tweets posted by the user. This includes the complete tweet object with all information related to tweet.
% \end{itemize}

\end{itemize}

\subsection{API Interface}
The full dataset is massive and the actual content cannot be directly distributed because of Twitter's policy~\footnote{https://developer.twitter.com/en/developer-terms/agreement-and-policy}. To help readers to better process the data, we have created an API~\footnote{https://github.com/KaiDMML/FakeNewsNet} that allows the users to download specific subsets of data. The API is provided in the form of multiple Python scripts which are well-documented and CSV file with news content URLs and associated tweet ids. In order
to initiate the API, the user must simply run the main.py
file with the required configuration. The API makes use of Twitter Access tokens fetch information related to tweets. Since FakeNewsNet includes multiple data sources, API provides options to select dataset of interest. Additionally, API facilitates user to download specific subsets of dataset like linguistic content only, visual content only, only  tweet information only, retweet information only, user information only and  social network only.
% {\color{blue} input , output}

\section{Potential Applications}\label{sec:application}
% {\color{blue} (may draw a figure to list potential fake news detection research directions and then state how the datasets can be used for these directions)}
FakeNewsNet contains information from multi-dimensions which could be useful for many applications. We believe FakeNewsNet would benefit the research community for studying various topics such as: (early) fake news detection, fake news evolution, fake news mitigation, malicious account detection.

\subsection{Fake News Detection}
One of the challenges for fake news detection is the lack of labeled benchmark dataset with reliable ground truth labels and comprehensive information space, based on which we can capture effective features and build models. FakeNewsNet can help the fake news detection task because it has reliable labels annotated by journalists and domain experts, and multi-dimension information from news content, social context, and spatiotemporal information.

First, news contents are the fundamental sources to find clues to differentiate fake news pieces. For example, a study has shown that the clickbait's headlines usually can serve as a good indicator for recognizing
fake news articles~\cite{chen2015misleading,shu2018clickbait}. In FakeNewsNet, we provide various attributes of news articles such as publishers, headlines, body texts, and videos. These information can be used to extract different linguistic features~\cite{hosseinimotlagh2018unsupervised} and visual features to further build detection models for clickbaits or fake news. For example,  style-based approaches try to detect fake news by capturing the manipulators in the writing style of news contents~\cite{potthast2017stylometric,wang2017liar}. In addition,  Knowledge-based approaches aim to use external sources to fact-check proposed claims in news content~\cite{shu2017fake}. Since we directly collect news articles from fact-checking websites such as PolitiFact and GossipCop, we provide the information of detail description and explanations from the fact-checkers, which are useful for us to learn common and specific perspectives of in what aspects the fake news pieces are formed.

%Studies \cite{shu2017exploiting,shu2017fake,wang2017liar,Mike2018} have shown that the linguistic features  have been widely used as an important feature for the detection task. FakeNewsNet has the linguistic features from the news source and help in the classification task.

Second, user engagements represent the news proliferation process over time, which provides useful auxiliary information to infer the veracity of news articles~\cite{kai2018social}.  Generally, there are three
major aspects of the social media context: users, generated posts, and networks. Since  fake news pieces are likely to be created and spread by non-human accounts, such as social bots or cyborgs~\cite{shao2017spread}. Thus, capturing users’ profiles and characteristics by user-based features can provide useful information for fake news detection. FakeNewsNet includes all the metadata for user profiles. In addition, people express their emotions or opinions towards fake news through social media posts, such as skeptical opinions, sensational reactions, etc. We collect all the user posts for the news pieces, as well as the second engagements (see Table~\ref{sec:related}) such as reposts, comments, likes, which can be utilized to extract abundant features, e.g., sentiment scores as in Figure~\ref{fig:user_feature_analysis}, to captures fake news patterns. Moreover, fake news dissemination processes tend to form an echo chamber cycle, highlighting the value of extracting network-based features to represent these types of network patterns for fake news detection. We provide a large-scale social network of all the users involving in the news dissemination process (see Table~\ref{sec:related}).

Third, early fake news detection aims to give early alerts of fake news during the dissemination process before it reaches a broad audience~\cite{liu2018early}.  Therefore early fake news detection methods are highly desirable and socially beneficial. For example, capturing the pattern of user engagements in the early phases could be helpful to achieve the goal of unsupervised detection~\cite{kai2019unsupervised}. Recent approaches utilize advanced deep generative models to generate synthetic user comments to help improve fake news detection performance~\cite{qian2018neural}.  FakeNewsNet contains all these types of information, which provides potentials to further explore early fake news detection models.

In addition, FakeNewsNet contains two datasets of different domains, i.e., political and entertainment, which can help to study common and different patterns for fake news under different topics. 

\subsection{Fake News Evolution}
% \textbf{Do we have time stamps of user comments, retweets, which can be used to trace how a fake news evolves}
% Studies show that the fake news evolve over its lifecycle. Using the timestamps of the user retweets and comments in the dataset one can trace how a fake news evolves. 
% {\color{blue}Write about different stages of fake news and how we can capture it}
The fake news diffusion process also has different stages in terms of people’s attention and reactions as time goes by, resulting in a unique life cycle. For example, breaking news and in-depth news demonstrate different life cycles in social media~\cite{castillo2014characterizing}, and social media reactions can help predict future visitation patterns of news pieces accurately even at an early stage. We can have a deeper understanding of how particular stories “go viral” from normal public discourse by studying the fake news evolution process. First, tracking the life cycle of fake news on social media requires recording essential trajectories of fake news diffusion in general~\cite{shao2016hoaxy}. Thus, FakeNewsNet has collected the related temporal user engagements which can keep track of these trajectories. Second, for a specific news event, the related topics may keep changing over time and be diverse for fake news and real news. FakeNewsNet is dynamically collecting associated user engagements and allows us to perform comparison analysis (e.g., see Figure~\ref{fig:dynamic_information}), and further investigate distinct temporal patterns to detect fake news~\cite{ruchansky2017csi}. Moreover, statistical time series models such as temporal point process  can be used to characterize different stages of user activities of news engagements~\cite{farajtabar2017fake}. FakeNewsNet enables the temporal modeling from real-world datasets, which is otherwise impossible from synthetic datasets.

\subsection{Fake News Mitigation}
Fake news mitigation aims to reduce the negative effects brought by fake news. During the spreading process of fake news, users play different roles such as \textit{provenances}: the sources or originators for publishing fake news pieces; \textit{persuaders}: who spread fake news with supporting opinions; and \textit{clarifiers}: who propose skeptically and opposing viewpoints towards fake news and try to clarify them. Identifying key users on social media is important to mitigate the effect of fake news. For example, the provenances can help answer questions such as whether the piece of news has been modified during its propagation. In addition, it’s necessary to identify influential persuaders to limit the spread scope of fake news by blocking the information flow from them to their followers on social media~\cite{fakebookchapter}. FakeNewsNet provides rich information about users who were posting, liking, commenting on fake news and real news pieces (see Figure~\ref{fig:ternary_plot}), which enables the exploration of identifying different types of users. 

To mitigate the effect of fake news, network intervention aims to develop strategies to control the widespread dissemination of fake news before it goes viral. Two major strategies of network intervention are: i) \textit{Influence Minimization}: minimizing the spread scope of fake news during dissemination process; ii) \textit{Mitigation Campaign}: Limiting the impact of fake news by maximizing the spread of true news. FakeNewsNet allows researchers to build a diffusion network of users with spatiotemporal information and thus can facilitate the deep understanding of minimizing the influence scopes. Furthermore, we may able to identify the fake news and real news pieces for a specific event from FakeNewsNet and study the effect of mitigation campaigns in real-world datasets.

\subsection{Malicious Account Detection}
Studies have shown that malicious accounts that can amplify the spread of fake news include social bots, trolls, and cyborg users.  Social bots are social media accounts that
are controlled by a computer algorithm. Social bots can give a false impression that information is highly popular and endorsed by many people, which enables the echo chamber effect for the propagation of fake news. 

We can study the nature of the user who spread fake news and identify the characteristics of the bot account used in the fake news diffusion process through FakeNewsNet. Using the feature like the user profile metadata and the historical tweets of the user who spread fake news along with the social network one could analyze the differences in characteristics of the users to clusters the users as malicious or not. Through a preliminary study in Figure~\ref{fig:bot_analysis}, we have shown that bot users  are more likely to exist in fake news spreading process. Even though existing work have studied the bot detection in general, but few studies investigate the influences of social bots for fake news spreading. FakeNewsNet could potentially facilitate the study of understanding the relationship between fake news and social bots, and further, explore the mutual benefits of studying fake news detection or bot detection.

\section{Conclusion and Future Work}\label{sec:conclude}
In this paper, we provide a comprehensive repository FakeNewsNet collected which contains information from news content, social context, and spatiotemporal information. We propose a principled strategy to collect relevant data from different sources. Moreover, we perform a preliminary exploration study on various features on FakeNewsNet and demonstrate its utility through fake news detection tasks over several state-of-the-art baselines. FakeNewsNet has the potential to facilitate many promising research directions such as fake news detection, mitigation, evolution, malicious account detection, etc.

There are several interesting options for future work. First, we will extend the FakeNewsNet repository to other reliable news sources such as other fact-checking websites or curated data collections. Second, we will improve the selection strategy used for web search results to reduce noise in the data collection process. We will also integrate FakeNewsNet repository with front-end software such as FakeNewsTracker~\cite{shu2018fakenewstracker}, and build an end-to-end system for fake news study.

\bibliographystyle{aaai}
\bibliography{refs}
\end{document}